# Anomaly of Film Porosity Dependence on Deposition Rate

Stephen P. Stagon and Hanchen Huang*
Department of Mechanical Engineering, University of Connecticut, Storrs, CT 06269

J. Kevin Baldwin and Amit Misra
MS K771, Los Alamos National Laboratory, Los Alamos, NM 87545

**Abstract**

This Letter reports an anomaly of film porosity dependence on deposition rate during physical vapor deposition – the porosity increases as deposition rate decreases. Using glancing angle deposition of Cu on $SiO_2$ substrate, the authors show that the Cu film consists of well separated nanorods when the deposition rate is 1 nm/second, and that the Cu films consists of a more uniform (or lower porosity) film when the deposition rate is 6 nm/second; all other deposition conditions remain the same. This anomaly is the result of interplay among substrate non-wetting, density of Cu nuclei on the substrate, and the minimum diameter of nanorods.

* The author to whom correspondence should be addressed; electronic mail: hanchen@uconn.edu

Metallic thin films of variable porosity are useful for a diverse range of applications. Examples of these applications are the use of Copper (Cu) nanorods in three-dimensional (3D) wafer bonding, metallic and bi-metallic nanorods as catalysts in chemical reactions, and nanorod coatings for enhanced boiling. 3D wafer bonding is an emerging technology which enables wafer level 3D integration of electronic systems [1]. Cu thin films have been used for wafer bonding, but wafer damage may occur at the high temperatures required to melt Cu thin films; bulk Cu has a melting temperature of 1085° C. Cu nanorods melt as low as 200° C, presenting an alternative to destructive high temperature processing [2, 3]. As catalysts, Pt nanorods have shown high catalytic activity due to their increased surface area when compared to bulk [4, 5]. Bi-metallic nanorods may also be made, such as Cu-Pt, which offer similar catalytic activity but with the use of less Pt, which has prohibitively high cost [5, 6]. In boiling applications, porous Cu nanorod films offer a significantly higher amount of boiling nucleation sites compared to micro and bulk films; up to a 30 fold increase [7]. In all applications, control of film porosity is desirable.

The control of film porosity is feasible through the competition of surface diffusion and geometrical shadowing during physical vapor deposition. Reducing surface diffusion and enhancing geometrical shadowing leads to higher film porosity [8]. Taking this approach to an extreme, researchers have grown nanorods using glancing or oblique angle deposition [9,10]. By geometrical shadowing, atomic flux from physical vapor lands on elevated surface regions. Further, by limiting the diffusion of adatoms, the deposited atoms do not reach depressed regions of the surface. As a result, elevated regions grow more elevated while depressed regions become more depressed as deposition progresses. According to this scenario, the porosity of thin films goes up as the deposition rate goes up, since the diffusion time of adatoms goes down with increasing deposition rate.

In this Letter, we report an anomaly in the dependence of film porosity on deposition rate. As the deposition rate goes up from 1 nm/second to 6 nm/second, the porosity goes down, and the film morphology changes from well-separated nanorods to a more uniform film. Further, we identify a mechanism for the anomaly, by considering the interplay of substrate non-wetting, nucleus density on the substrate, and the minimum diameter of nanorods.

Before presenting the results, we briefly describe the experimental methods of growth and characterization. Cu films are deposited on $SiO_2$ - or a layer of native oxide on top of Si(111) - substrates using electron-beam evaporation. The vacuum chamber, a cylinder of roughly 50 cm diameter and 75 cm tall, is first evacuated to a base pressure of $1 \times 10^{-4}\ Pa$ and held for several hours to remove impurities from the chamber. During deposition, working pressure is about $4 \times 10^{-4}\ Pa$. The substrate and chamber are not temperature controlled, and deposition occurs with chamber temperatures ranging from 20° C to 45° C. The source materials is 99.997% copper and is installed in the vacuum chamber before deposition. The substrate is 40 cm away from the source. To achieve geometrical shadowing, the substrate is oriented approximately $85°$ from the top of the chamber, which is parallel to the source. Source materials are then vaporized using a high power electron beam arc. Varying deposition rate from 0.1 nm/second to 6 nm/second, films are grown to thicknesses up to 500 nm. The deposition rate is measured perpendicular to the flux using a quartz crystal deposition monitor. Due to the significant angle of inclination, the actual deposition rate is lower than measured. Samples are removed from vacuum and characterized using a FEI Quanta 250 FEG field emission scanning electron microscope. Crystallographic orientation is characterized using a Bruker DB8 Advance Diffractometer System.

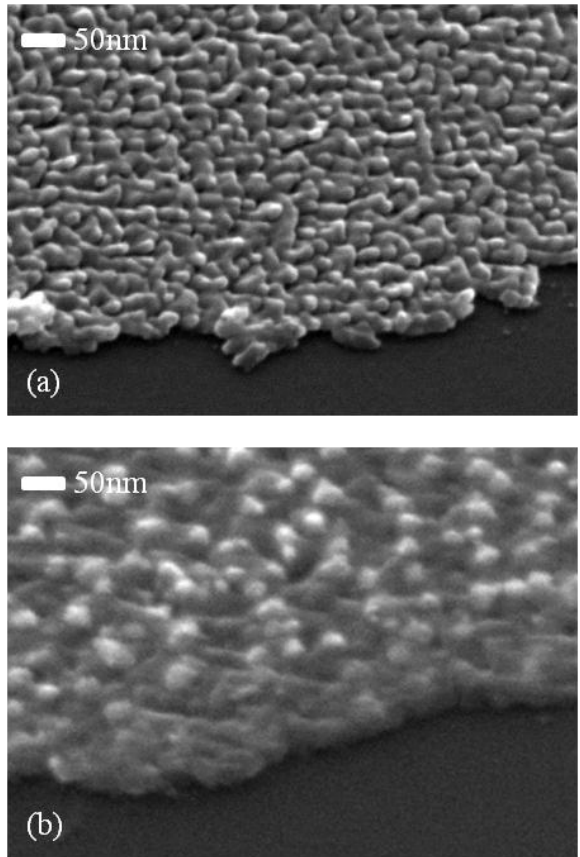

Fig. 1: Scanning electron microscopy (SEM) images of Cu films deposited at (a) 1 nm/second and (b) 6 nm/ second.

As shown in Fig. 1, the Cu film has large porosity at the deposition rate of 1 nm/second, and separated nanorods are identifiable. As the deposition rate increases from 1 nm/second to 6 nm/second, the film porosity decreases and a more uniform film develops. The decrease of porosity with increasing deposition rate is anomalous with respect to the conventional understanding.

We suggest a mechanism for the anomaly, based on the interplay of substrate non-wetting, nucleus density, and the minimum diameter of nanorods. First, due to non-wetting of Cu on $SiO_2$, according to literature reports [11], 3D nuclei of Cu form at the early stage. Second, a higher deposition rate leads to a higher density of nuclei, and forces a complete coverage of the non-wetting substrate before nanorods develop. The coverage of a non-wetting barrier layer by Cu in the metallization of integrated circuits is based on the same principle [12]. Third, as the separation of nuclei is smaller than the minimum diameter of Cu nanorods - which is on the order of 30 nm, according to [13] under the condition that surface steps are always multiple-layer - the porosity between nuclei disappears and a uniform film develops. Based on recent studies, multiple-layer steps are kinetically stable [14], in contrast to earlier understanding [15]. When the initial diameter of an island is smaller than this minimum, the overwhelming dominance of multiple-layer step is feasible as the island expands laterally from bottom to top; this is the condition provided by the non-wetting condition.

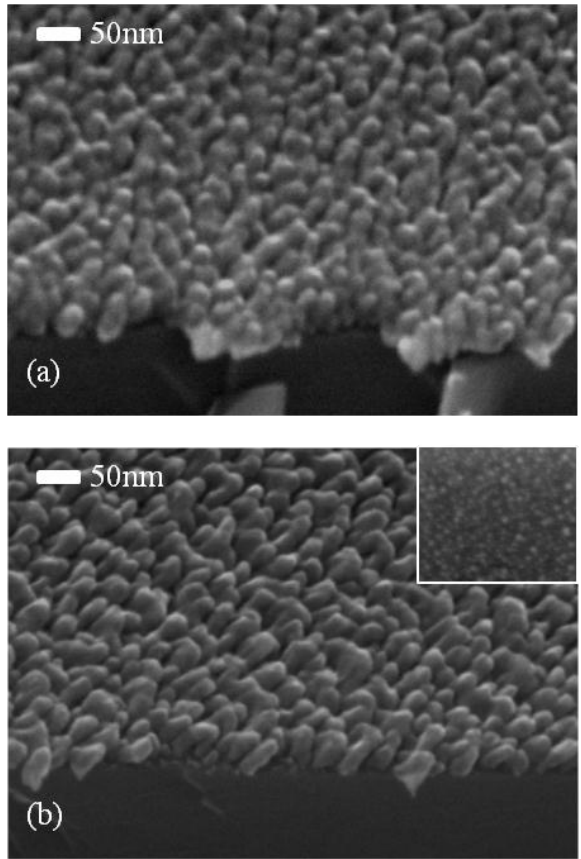

Fig. 2: Scanning electron microscopy (SEM) images of Cu films deposited at 0.1 nm/second, (a) without indium and (b) with 5 nm indium pre-deposition; with the inset being an SEM image of the indium nuclei on a area of 500 nm X 500 nm of $SiO_2$.

To validate the suggested mechanism, we have designed two experiments. In the first experiment, we use a deposition rate of 0.1 nm/second, which is 10X lower than that of Fig. 1(a). The critical deposition rate that separates a large-porosity film and a uniform film is between 1 and 6 nm/second. Further lowering the deposition rate from 1 nm/second to 0.1 nm/second should also lead to well-separated nuclei and thereby nanorods, according to our suggested mechanism. As shown in Fig. 2(a), the film consists of nanorods that are similar to those in Fig. 1(a). This similarity indicates that the suggested mechanism is valid. In the second experiment, we pre-deposit indium (In) of 5 nm in thickness on the $SiO_2$ substrate, before Cu deposition. 3D nuclei of In on the substrate (as shown in the inset of Fig. 2(b)) serve as preferential nucleation sites for Cu, due to the strong In-Cu interaction [16]. This heterogeneous nucleation of Cu islands further promotes their separation, and should lead to better separated nanorods than without In pre-deposition, if the suggested mechanism is valid. As shown in Fig. 2(b), the Cu nanorods indeed are better separated than those in Fig. 2(a).

Having presented the anomaly and a feasible mechanism, we next examine how the films will develop further as deposition continues, and what texture dominates. As shown in Fig. 3(a), the large-porosity film develops into nanorods of very small diameter, on the order of 30 nm, that is comparable to the minimum diameter of Cu nanorods [13]. From the uniform film of Fig. 1(b), nanorods still develop as deposition continues, due to the geometrical shadowing at glancing angle incidence. The density of Cu nanorod nuclei on the film surface of Fig. 1(b) derives from growth of Cu on Cu, and non-wetting disappears. As a result, the nuclei of Cu nanorods are not bounded by overwhelmingly multiple-layer steps. Since the realization of the minimum diameter of Cu nanorods depends on the dominance of multiple-layer steps, the diameter of Cu nanorods is larger than the minimum. As an indication of out-of-plane texture, the x-ray diffraction (XRD) of Fig. 4 shows that both films of Fig. 3 have the <111> texture, as most Cu films do [17, 18].

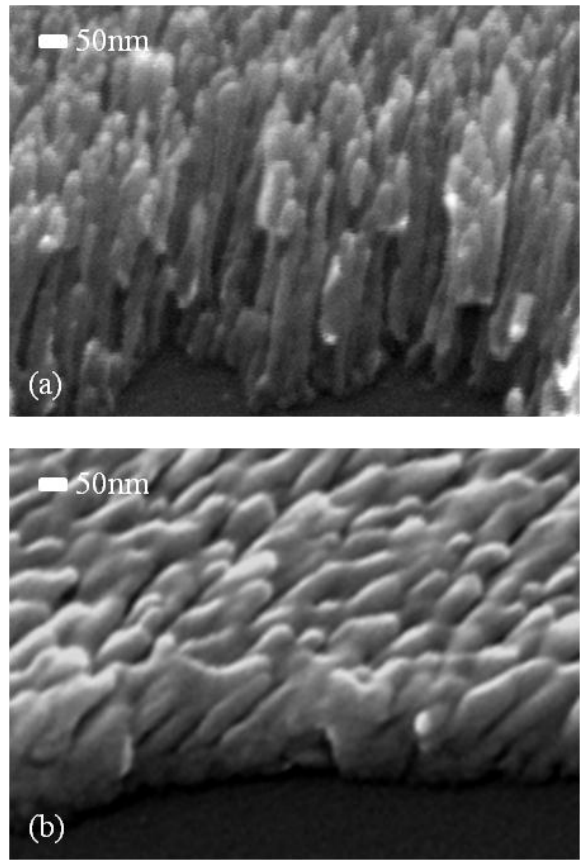

Fig. 3: Scanning electron microscopy (SEM) images of Cu films deposited at (a) 1 nm/second and (b) 6 nm/ second.

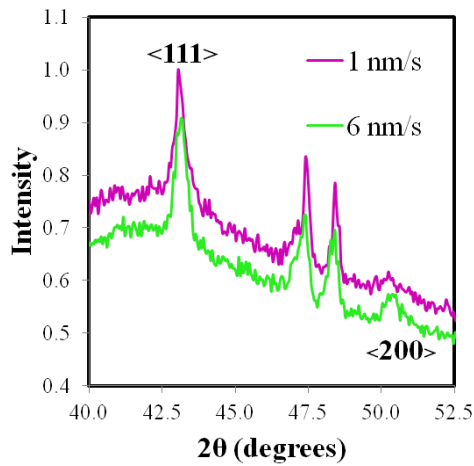

Fig. 4: x-ray diffraction (XRD) of the two films in Fig. 3.

In summary, we report an anomaly in the dependence of film porosity on deposition rate, suggest a mechanism for the anomaly, and provide two pieces of evidence to validate the suggested mechanism. On a non-wetting substrate, 3D nuclei form at the early stage, and geometrical shadowing at a glancing angle promotes further 3D growth leading to porous films. As the separation of nuclei is larger than the minimum diameter of nanorods, well-separated nanorods develop and porosity is large. However, when the separation is smaller than the minimum diameter, the nanorods will not have space to develop, uniform film develops and the porosity is low.

**Acknowledgement:** The authors acknowledge financial support from the Department of Energy Office of Basic Energy Sciences (DE-FG02-09ER46562), and the provision of facilities at Center for Integrated NanoTechnologies at Los Alamos and Sandia National Laboratories. HH also acknowledges financial support of the National Science Foundation (CMMI-0856426) in bridging the growth of thin films with that of nanorods.